\documentclass[%
,twocolumn
,secnumarabic%
,amssymb, amsmath,nobibnotes, aps, prl]{revtex4}

\usepackage{graphicx}
\usepackage[usenames]{color}

\begin{document}

\title{Tuning the Mass of Chameleon Fields in Casimir Force Experiments}

\author{Ph.~Brax}
\email{philippe.brax@cea.fr}
\affiliation{Institut de Physique Th\'{e}orique, CEA, IPhT, CNRS, URA 2306, F-91191Gif/Yvette
Cedex, France}
\author{C.~van~de~Bruck}
\email{c.vandebruck@shef.ac.uk}
\affiliation{Department of Applied Mathematics, University of Sheffield Hounsfield Road,
Sheffield S3 7RH, United Kingdom}
\author{A.~C.~Davis}
\email{a.c.davis@damtp.cam.ac.uk}
\affiliation{Department of Applied Mathematics and Theoretical Physics, Centre for Mathematical
Sciences, Cambridge CB3 0WA, United Kingdom}
\author{D.~J.~Shaw}
\email{d.shaw@qmul.ac.uk}
\affiliation{Queen Mary University of London, Astronomy Unit, Mile End Road, London E1 4NS, United
Kingdom}
\author{D.~Iannuzzi}
\email{iannuzzi@few.vu.nl} \affiliation{Department of Physics and Astronomy, VU University
Amsterdam, De Boelelaan 1081, 1081 HV Amsterdam, The Netherlands}

\date{\today}

\begin{abstract}
We have calculated the chameleon pressure between two parallel plates in the presence of an
intervening medium that affects the mass of the chameleon field. As intuitively expected, the gas
in the gap weakens the chameleon interaction mechanism with a screening effect that increases with
the plate separation and with the density of the intervening medium. This phenomenon might open up
new directions in the search of chameleon particles with future long range Casimir force
experiments.
\end{abstract}

\maketitle

In 2004, J. Khoury and A. Weltman showed that a scalar field whose mass depends on the local matter
density may explain the late-time acceleration of the Universe and still elude laboratory tests of
the gravitational inverse-square law~\cite{Khoury,Brax1,Mota,Brax2}. On cosmological scale, where
the local density is small, the mass of the field is sufficiently low to drive the expansion of the
Universe. In table-top gravity experiments, however, the mass of the field is largely increased by
the higher value of the local matter density, and the interaction mediated by the field becomes too
small to be detected. Because of their ability to adapt to the surroundings, those fields are
generally known as \textit{chameleon fields}.

Chameleon particles have never been detected in any laboratory experiment. Still, one can show that
the chameleon pressure between two plates kept parallel in vacuum at a separation of a few tens of
$\mu$m is of the same order of magnitude as the Casimir attraction under the same
conditions~\cite{Brax2}, reaching values that are within the detection sensitivity of the next
generation of long range (i.e., separations much larger than 1 $\mu$m) Casimir force
experiments~\cite{onofrio1}. Unfortunately, long range Casimir force experiments suffer from one
main drawback. Since the Casimir pressure rapidly decreases with increasing separation, large
surfaces (several hundreds of cm$^2$) are needed to reach the force detection limit. The
electrostatic potential of a large surface, however, is typically non-uniform, giving rise to
background forces that easily overcome the Casimir
attraction~\cite{long1,long2,onofrio2,carugno,lamoreaux}. Those electrostatic forces are difficult
to control and to quantify independently, making accurate Casimir force experiments at large
separation a challenge that has still to be solved. Without reliable Casimir force measurements, it
is of course not possible to use Casimir force set-ups to assess the existence of chameleon fields.

In this letter we propose a novel approach that might alleviate the problem described above and open
new possibilities for the detection of chameleon fields in long range Casimir force measurements.
The idea is to measure the total force between two parallel plates as a function of the density of
a neutral gas allowed into the cavity. As the density of the gas increases, the mass of the
chameleon field in the cavity increases, giving rise to a screening effect of the chameleon
interaction. If all the other relevant forces between the two plates (Casimir and electrostatic) do
not depend on the density of the gas in the gap, a direct comparison of the results obtained at low
densities (strong chameleon force) with those obtained at high densities (weak chameleon force)
should allow one to detect or to rule out the existence of chameleon particles.

To demonstrate the concept behind our proposal, we have calculated the expected chameleon pressure
between two parallel plates at separation $d$ immersed in a gaseous atmosphere of density $\rho$.
To illustrate the chameleon effect, we have focused on chameleon potentials of the form
\begin{equation}
V(\phi)= \Lambda^4 + \frac{\Lambda^{4+n}}{\phi^n},
\end{equation}
where $\Lambda\approx 2.4\cdot 10^{-12}$ GeV is the energy density leading to the  late time
acceleration of the universe. The chameleons are also coupled to matter, and the effective
potential in the presence of matter is
\begin{equation}
V_{\rm eff}(\phi)=V(\phi) + \rho e^{\beta \phi/m_{\rm Pl}},
\end{equation}
where $\beta$ is the coupling constant and $m_{\rm Pl}\approx 2\cdot 10^{18}$ GeV is the reduced
Planck mass. The matter density $\rho$ influences the shape of the effective potential leading to
an effective minimum. In particular, the mass, $m$, of the chameleon field at the minimum satisfying
\begin{equation}
\partial_\phi V(\phi_{\rm min})= -\frac{\beta \rho}{m_{\rm Pl}},
\label{phi}
\end{equation}
with $\phi_{\rm min}\ll m_{\rm Pl}$ in the situations of interest, becomes density dependent
\begin{equation}
m^2= \frac{\beta \rho}{m_{\rm Pl}}\left( \frac{n+1}{\phi_{\rm min}} + \frac{\beta}{m_{\rm Pl}}\right).
\label{m}
\end{equation}
Let us now consider the Casimir set-up with two parallel plates. We will denote by $m_b$ and
$\phi_b$ the value of the chameleon mass and field in the vacuum corresponding to the matter
density $\rho$ in between the plates and given by eq. (\ref{m}) and eq. (\ref{phi}). When the plates
are present, the chameleon field in between the plates is not the vacuum one. It acquires a profile
with a minimum $\phi_0$ at the midpoint between the two plates. Denoting $n=(1-p)/p$ and
$z=(\phi_0/\phi_b)^{1/p}$, we find that the value of the field $\phi_0$ and the distance are
related as
\begin{equation}
d= \frac{\sqrt 2 z^{(1+p)/2}}{m_b}\int_0^1 \frac{x^{p-1}dx}{\sqrt{h_{p-1}(x)- z h_p(x)}},
\end{equation}
where $h_p(x)=(1-x^p)/p$ and the pressure is
\begin{equation}
\frac{F_\phi}{A}=\frac{n+1}{n} \frac{\Lambda^{4+n}}{\phi_b^n} z^{p-1} [h_{1-p}(z)-z h_{-p}(z)].
\label{f_cham}
\end{equation}
This parametric representation allows one to calculate the pressure profile as a function of $d$.
When $m_c^{-1}\ll d\ll m_{b}^{-1}$ where $m_c$ is the chameleon mass inside the plates, we find an
algebraic decay
\begin{equation}
\frac{F_\phi}{A}=O(\Lambda^4(\Lambda d)^{2(p-1)/(p+1)}).
\end{equation}
Notice that the energy scale $\Lambda$ corresponds to a length scale $\Lambda^{-1} \approx 82 \mu
{\rm m}$. This explains why the chameleonic pressure comes to the fore when distances are on the
order of 10 $\mu \rm m$ and more. When the matter density in between the plates increases, the
distance satisfies $d\gg m_b^{-1}$ as the mass $m_b$ increases. In this regime, the chameleonic
pressure behaves like $\exp (-m_b d)$ and is therefore exponentially suppressed. Hence, by
increasing the density between the plates at a fixed distance $d$, one would observe a contribution
from the chameleon to the pressure on the plates. When the density increases, the chameleon
pressure is screened off due to the exponential fall-off resulting from the large chameleon mass
$m_b$.

\begin{figure}
\includegraphics[scale=0.3,angle=-90]{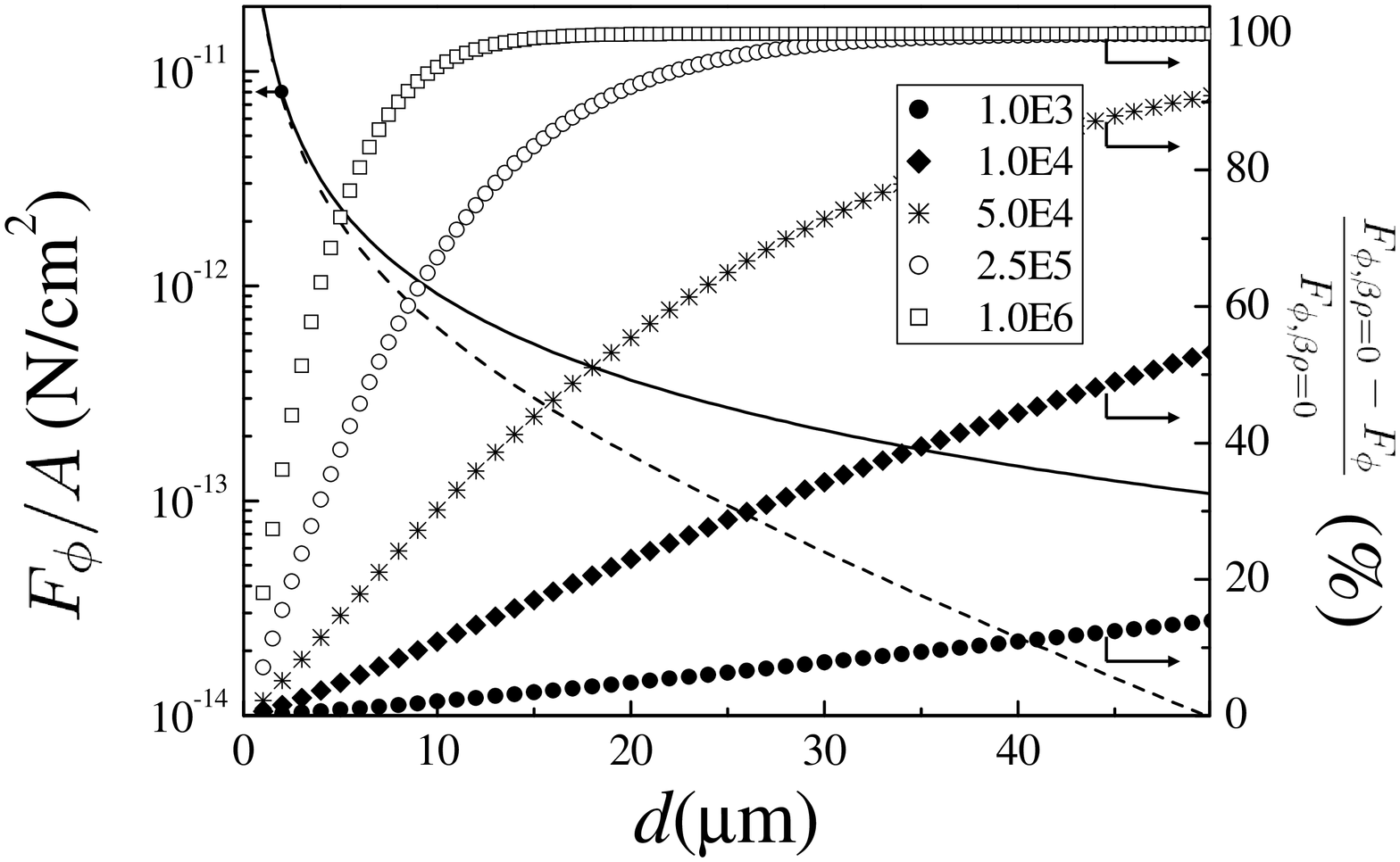}
\caption{Continuous and dashed lines (left y-axis): Chameleon pressure between two parallel plates
as a function of separation for two configurations: in vacuum (continuous line) and in a gas with
density $\rho\approx 5$ g/l. The calculations were performed fixing the exponent that describes the
chameleon potential to $n=4$ and the coupling constant to $\beta=10000$. Symbols (right y-axis):
Decrease of the chameleon pressure (in percentage) between two parallel plates induced by the screening effect, plotted as a function of separation. The different symbols correspond to the different values of $\beta \rho$ indicated in
the legend, where $\rho$ is expressed in g/l.} \label{fig_pressure}
\end{figure}

\begin{figure}
\includegraphics[scale=0.3,angle=-90]{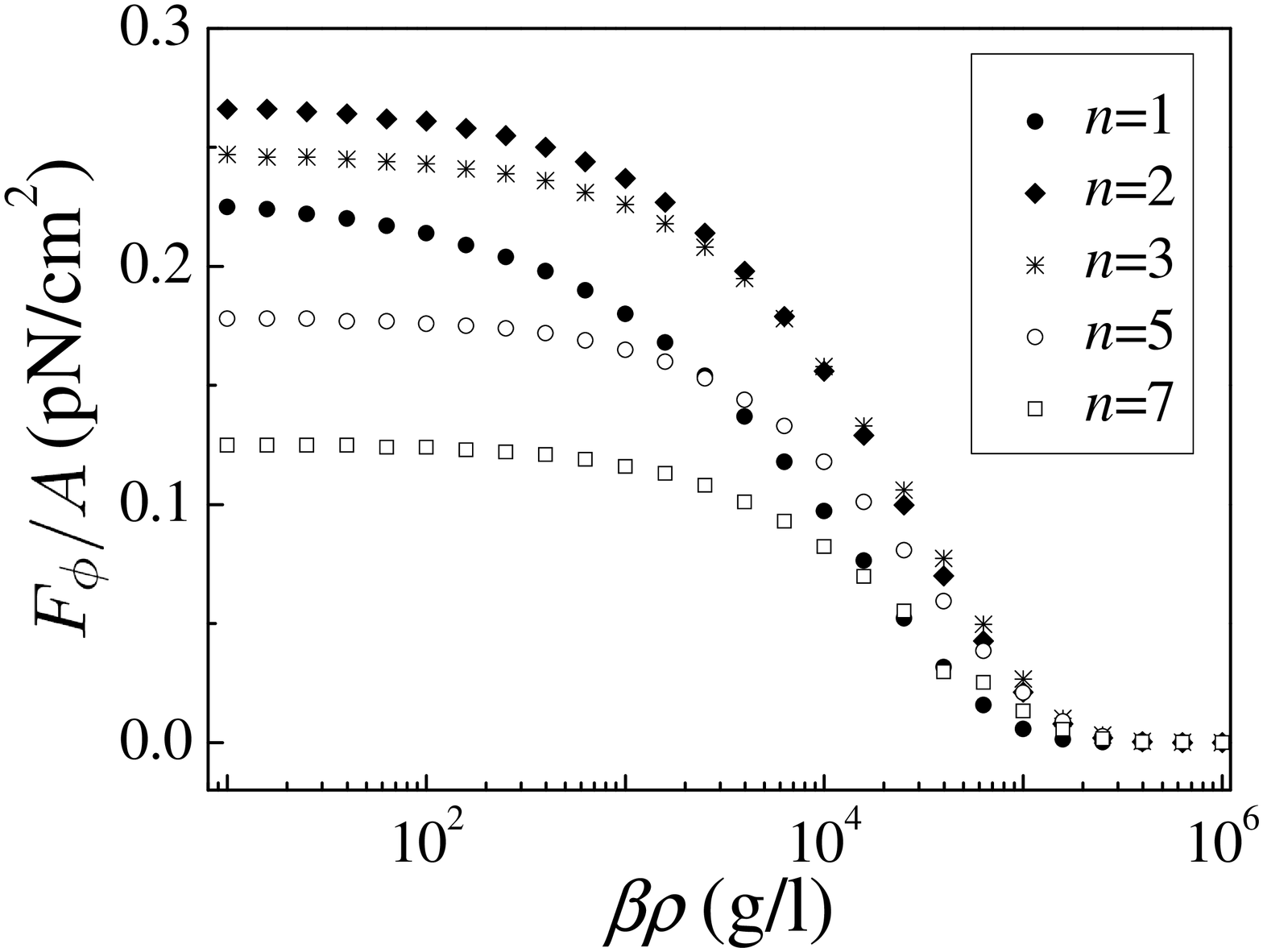}
\caption{Expected chameleon pressure between two parallel plates kept at 30 $\mu$m separation as a
function of $\beta \rho$, where $\beta$ is the coupling constant and $\rho$ is the density of the
gas in the gap, expressed in g/l. Different symbols correspond to different values of the exponent
$n$ that describes the chameleon potential.} \label{fig_bro_30}
\end{figure}

This behavior is well illustrated in Fig. \ref{fig_pressure}. As an example, we fixed $n=4$ and
$\beta=10000$. The continuous and the dashed lines (left y-axis) represent the calculated chameleon
pressure in vacuum and in a gas with $\rho\approx 5$ g/l, respectively. The symbols (right y-axis)
show the decrease of the chameleon pressure (in percentage) induced by the screening effect for
different values of $\beta \rho$ (expressed in g/l) and for $n=4$. As intuition suggests, the
screening effect is stronger at larger separations and for larger values of $\beta \rho$.
Importantly, room temperature, atmospheric pressure gases can indeed give rise to large screening
effects already at separations where long range Casimir force set-ups are designed to work. Similar
behaviors are expected for different values of $n$, as illustrated in Fig. \ref{fig_bro_30}, where
we report the calculated chameleon pressure between two parallel plates kept in a gaseous
atmosphere of density $\rho$ as a function of $\beta \rho$ for $d=30$ $\mu$m.

Let us now consider an experimental set-up designed to measure the total interaction between two
gold-coated parallel plates kept at $d=30$ $\mu$m in  vacuum. At that separation, the chameleon
force is of the same order of magnitude as the Casimir force, reaching approximately 0.2 pN/cm$^2$
-- a value that appears to be within the sensitivity of future long-range Casimir force
experiments~\cite{onofrio1}. Unfortunately, both interactions mechanisms are expected to be largely
dominated by the electrostatic interaction due to the non-uniformity of the surface potential.
Following \cite{speake} and \cite{carugno}, in fact, it is possible to estimate the electrostatic
pressure as:

\begin{equation}
\frac{F_{el}}{A}=\epsilon_0 \left(
\frac{\sigma_L^2}{2d^2}+\frac{2\sigma_S^2}{k_{max}^2-k_{min}^2}\int_{k_{min}}^{k_{max}}\frac{k^3}{\sinh^2(kd)}dk
\right), \label{f_el}
\end{equation}
where $\epsilon_0$ is the dielectric permittivity of vacuum, $\sigma_L$ and $\sigma_S$ are the
variances of long and short wavelength components of the surface potential, and
$k_{min,max}=2\pi/\lambda_{max,min}$, with $\lambda_{max,min}$ representing the maximum and minimum
characteristic sizes of the short range variations of the surface potential. Using a tentative value of
$\sigma_{L,S}\simeq 50$ mV \cite{carugno} to evaluate the first term and adapting the approach
described in \cite{speake} to evaluate the second term, one obtains $F_{el}/A\approx 2 \cdot 10^3$
pN/cm$^2$, i.e., 4 orders of magnitude larger than the Casimir and the chameleon pressure. Although
this value represents an upper limit, it inarguably shows how electrostatic forces can easily
dominate in long range Casimir or chameleon force experiments conducted in vacuum.

Let us  however  suppose that the experiment is performed in a gaseous atmosphere, and let us
analyze how the presence of the gas in the gap changes the chameleon, Casimir, and electrostatic
interactions. By way of example, we will assume that the gap is filled with Xe at room temperature
($T=293.15$ K) and at a pressure $P$ that can be varied up to 0.5 atm. In this range of pressures,
the density of the gas is well approximated by: $\rho=5.462\cdot P$ g/l~\cite{lemmon}, and its
dielectric constant can be described by the limit for low densities of the Lorentz-Lorenz equation:
$\epsilon=\epsilon_0+N\alpha$, where $N$ is the density expressed in atoms/m$^3$ and $\alpha=4
\cdot 10^{-40}$ Fm$^2$ is the atomic polarizability of Xe~\cite{mairland}.

\begin{figure}
\includegraphics[scale=0.3,angle=-90]{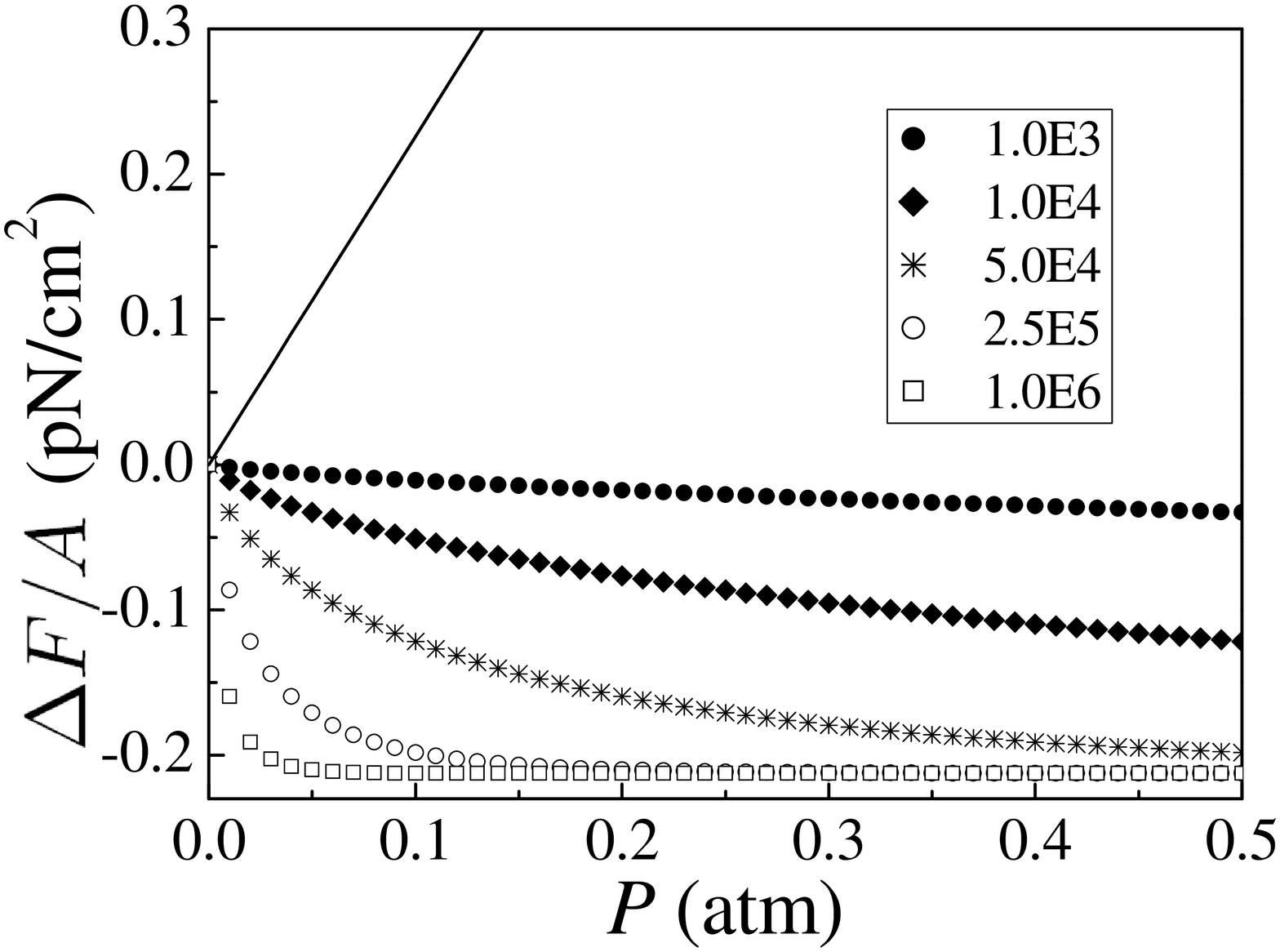}
\caption{Symbols: Expected change of the chameleon pressure between two parallel plates kept at 30
$\mu$m separation when the gap is filled with gaseous Xe at room temperature and pressure $P$.
Different symbols correspond to different values of the coupling constant $\beta$. Line: expected
change of the electrostatic force in the same experimental configuration.} \label{fig_comparison}
\end{figure}

\noindent \textit{Chameleon pressure:} because of the screening effect previously described, the
chameleon attraction is expected to \textit{decrease} as the pressure of the gas increases. Using
eq. \ref{f_cham} with, for example, $n=4$, one obtains the curves reported in Fig.
\ref{fig_comparison}. It is evident that this range of pressures is already sufficient to reduce
the chameleon attraction of $\simeq 0.1$ pN/cm$^2$ as soon as $\beta>10000$.

\noindent \textit{Casimir pressure:} Because of the increase of the dielectric function of the
intervening medium, the Casimir attraction is expected to \textit{decrease} as the pressure of the
gas increases. The effect is however extremely small. In first approximation (large separations and
$T\rightarrow 0$), in fact, the Casimir force between two plates scales like
$1/\sqrt{\epsilon}$~\cite{parsegian}, corresponding to a maximum decrease of $\approx 0.025\%$ at
$P=0.5$~atm.

\noindent \textit{Electrostatic pressure:} Because of the increase of the dielectric constant of
the intervening medium, the electrostatic attraction is expected to \textit{increase} linearly with
the pressure of the gas. The electrostatic force thus follows a behavior that is opposite to what
is expected for the chameleon force. In terms of absolute values, taking the upper limit of
$F_{el}/A\approx 2 \cdot 10^3$ pN/cm$^2$ as an estimate of the electrostatic force in vacuum, one
can calculate that the presence of the gas in the gap increases the electrostatic force of an
amount that, at least at lower pressures, is of the same order of magnitude as the change on the
chameleon force as soon as $\beta>10000$, as illustrated in Fig. \ref{fig_comparison}.

\begin{figure}
\includegraphics[scale=0.3,angle=-90]{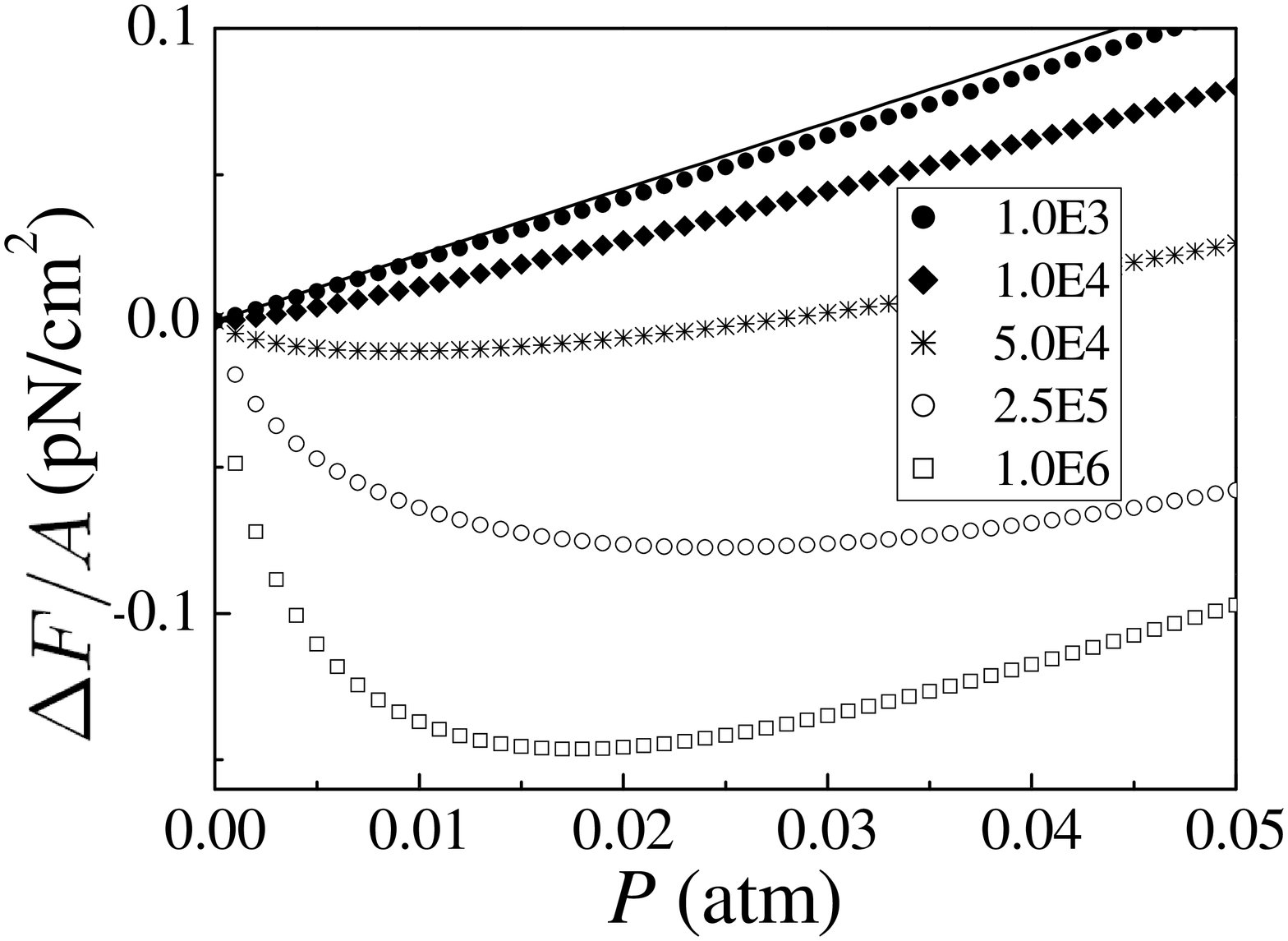}
\caption{Expected change of the total pressure between two parallel plates kept at 30
$\mu$m separation when the gap is filled with gaseous Xe at room temperature and pressure $P$.
Different symbols correspond to different values of the coupling constant $\beta$. The line represents the result expected in the absence of chameleon interaction.} \label{fig_exp}
\end{figure}

We conclude that, while in a force-vs-distance experiment in vacuum chameleon fields manifest themselves as a $\simeq 0.1$~pN/cm$^2$ pressure compared to  a background electrostatic pressure of $\simeq 10^3$~pN/cm$^2$, a force-vs-density experiment would only need to distinguish anomalies of the total interaction strength of $\simeq 0.01$~pN/cm$^2$ or larger compared to  a $\simeq 0.1$~pN/cm$^2$ background (see Fig. \ref{fig_exp}). A force-vs-density experiment can thus alleviate the problem of the electrostatic force of at least 3 orders of magnitude. Furthermore, in a force-vs-density experiment the electrostatic and
the chameleon forces are supposed to manifest different behaviors: the electrostatic force, in
fact, increases with gas density, while the chameleon force is expected to decrease. The approach
described in this letter might thus open up new directions in the search for chameleon particles,
although it is fair to stress that the practical implementation of this kind of experiments is not
straightforward. We would like to point out, for example, that $\sigma_{L,S}$ must remain constant
within a few $\mu$V to avoid that the effect of the variations in the non-uniformity of the surface
potential overcomes the expected change of the chameleon pressure (see eq. \ref{f_el}). Similarly, for $\sigma_{L,S}\simeq 50$ mV, the separation between the two surfaces must remain constant within less than $1$ nm to keep the errors due to variations in $d$ lower than the signal that has to be detected. This constraint, however, would be less stringent for smaller values of $\sigma_{L,S}$.

In conclusion, we have calculated the chameleon force between two parallel plates in the presence
of an intervening medium. According to our calculations, a gas with a density of the order of a few
g/l increases the mass of the chameleon so effectively that, at separations of the order of a few
tens of $\mu$m, the chameleon interaction is significantly screened out. We suggest that this
mechanism might be used to unravel the existence of chameleon fields in future long range Casimir
force set-ups, where one could measure the chameleon force between two parallel plates as an
anomalous dependence of the total interaction strength on the density of a gas allowed in the gap.

This work was partially stimulated by interesting discussions within the ESF Research Network
CASIMIR, and in particular with S. de Man, K. Heeck, G. Ruoso, G. Bimonte, G. Carugno, C. Speake, A. F.
Borghesani, and R. Onofrio. DI acknowledges support from the European Research
Council under the European Community's Seventh Framework Programme (FP7/2007-2013)/ERC grant
agreement 201739 and from the Netherlands Organisation for Scientific Research (NWO), under the
Innovational Research Incentives Scheme VIDI-680-47-209. DJS is supported by STFC and
the work of CvdB and ACD is, in part, supported by STFC.

\end{document}